# Exploring the Performance Benefit of Hybrid Memory System on HPC Environments

Ivy Bo Peng*, Roberto Gioiosa‡, Gokcen Kestor‡, Pietro Cicotti†, Erwin Laure* and Stefano Markidis*
*Department of Computational Science and Technology, KTH Royal Institute of Technology, Sweden
†Advanced Technology Laboratory, San Diego Supercomputer Center, USA
‡Computational Science and Mathematics Division, Pacific Northwest National Laboratory, USA

***Abstract*—Hardware accelerators have become a de-facto standard to achieve high performance on current supercomputers and there are indications that this trend will increase in the future. Modern accelerators feature high-bandwidth memory next to the computing cores. For example, the Intel Knights Landing (KNL) processor is equipped with 16 GB of high-bandwidth memory (HBM) that works together with conventional DRAM memory. Theoretically, HBM can provide $\sim 4\times$ higher bandwidth than conventional DRAM. However, many factors impact the effective performance achieved by applications, including the application memory access pattern, the problem size, the threading level and the actual memory configuration. In this paper, we analyze the Intel KNL system and quantify the impact of the most important factors on the application performance by using a set of applications that are representative of scientific and data-analytics workloads. Our results show that applications with regular memory access benefit from MCDRAM, achieving up to $3\times$ performance when compared to the performance obtained using only DRAM. On the contrary, applications with random memory access pattern are latency-bound and may suffer from performance degradation when using only MCDRAM. For those applications, the use of additional hardware threads may help hide latency and achieve higher aggregated bandwidth when using HBM.**

## I. Introduction

Current and next-generation supercomputers will heavily rely on hardware accelerators to achieve extreme parallelism and high performance. Accelerators come in many forms, including general-purpose GPUs, FPGAs, or other energy-efficient architectures, such as ARM big.LITTLE [1] and Intel Many Integrated Core (MIC) [2]. Accelerators generally present a different architecture compared to the host processors and require programmers to adapt their applications to fully exploit the underlying hardware features. Besides a large number of computing threads, modern accelerators, such as AMD Fiji, NVIDIA Pascal GPUs, and Intel Knights Landing (KNL), also feature a relatively small, high-bandwidth on-chip memory implemented with 3D-stacked technology. This memory can be used to cache application data but also as an extension of DRAM. The effective use of these memory technologies helps reducing data movement [3] and is of paramount importance to achieve high performance on modern and future supercomputers.

The simultaneous presence of two different memory technologies that work side-by-side, e.g., the traditional DRAM and the new on-chip memory, makes the system heterogeneous from the memory perspective. Deeper and heterogeneous memory systems are likely to be the norm in future mainstream HPC systems. Thus it is important to understand how applications can make efficient use of these new high-bandwidth memory technologies and what the limitations and constraints are [4]. While several studies have successfully ported specific applications to hardware accelerators equipped with 3D-stacked memory [5], [6], a deep and comprehensive understanding on how to leverage the high-bandwidth memory in HPC applications is still an open issue. In this paper we target this problem and aim at providing recommendations to programmers on which applications may benefit from these new memory technologies and what are the limiting factors to consider and their impact on the application performance. Without loss of generality, we analyze the performance benefits of the new 3D-stacked high-bandwidth memory (HBM) packaged with the Intel KNL processor, the second-generation Intel Xeon Phi processor. This processor provides the main computing power of the Cori [7] supercomputer at NERSC, while the upcoming Aurora supercomputer at Argonne National Laboratory [8] will use the third-generation Intel Xeon Phi processors.

The HBM 3D-stacked memory on the KNL processor is implemented as a Multi-Channel DRAM (MCDRAM). DRAM and MCDRAM differ significantly in the metrics of capacity, bandwidth and latency. In addition, Intel provides three memory modes to configure HBM: *flat*, *cache*, and *hybrid* mode. In flat mode, the user sees HBM as a memory pool exposed by the operating system (OS) as an additional NUMA node. In cache mode, HBM is transparent to the OS and is directly managed by the hardware as a large cache for DRAM transfers. This is the expected default mode, as it requires neither modifications to the application code nor re-structuring of the algorithm or data structures. In hybrid mode, a part of HBM is explicitly managed by the user (as in the flat mode), while the rest is managed by the hardware as a cache for DRAM transfers. Because of the capacity, bandwidth, and latency differences between DRAM and HBM, and also because of the possibility of configuring the memory system in different modes, performance will vary widely depending on the application characteristics, the usage of memory resources, and the chosen configuration. The goal of this work is to investigate the correlation between application characteristics and attainable performance on Intel KNL processor.

We follow a two-step methodology. First, we characterize the memory performance on KNL systems through precise micro-benchmarks. From the extracted hardware characteristics, such as memory bandwidth and latency, we identify factors correlated to the observed performance. Next, we perform an exhaustive performance evaluation using kernels (DGEMM and GUPS), two scientific DOE proxy applications (MiniFE and XSBench), and one benchmark mimicking data analytics workloads (Graph500). In this step we vary different parameters to collect performance measurements from a vast array of configurations including different problem sizes, number of OpenMP threads, and memory modes (i.e., flat or cache). Analyzing the results we were able to assess performance variations due to different configurations. Thus we can evaluate the application sensitivity to the memory configurations, including a quantification of the performance improvement/degradation due to memory configuration and usage.

Our results show that the memory access pattern impacts the application performance the most. Applications with regular data access, such as DGEMM and MiniFE, are mostly limited by memory bandwidth, thus they considerably benefit from HBM. These applications can achieve $3\times$ performance improvement when using HBM compared to solely using DRAM. However, when the problem size is larger than the HBM capacity, the benefit of using HBM as cache decreases with increasing problem size. The performance of applications with random memory access, such as GUPS, Graph500 and XSBench, is largely affected by the memory latency and does not benefit from using HBM, as the latency is higher than the DRAM. We also found that the use of additional hardware threads per core leads to a performance improvement when using HBM as the simultaneous use of multiple hardware threads can help hide HBM latency. For MiniFE, we observe a $3.8\times$ performance improvement with respect to the performance obtained with only DRAM when we use four hardware threads per core.

In this paper we make the following contributions:

1) We identify three main factors (access pattern, problem size, and threading) that determine application performance on systems with KNL processors.
2) We quantify the impact of using different memory configurations (HBM-only, cache and DRAM-only), problem size, number of OpenMP threads on application performance.

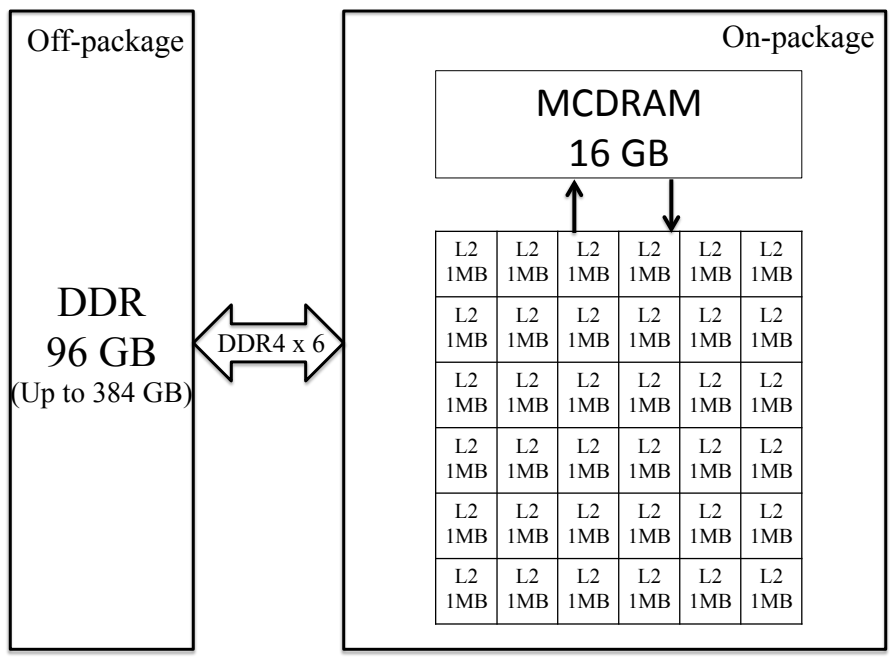

Figure 1: Layout of different memories on KNL. The grid indicates the Intel mesh of tiles. Each tile consists of two cores sharing a 1-MB L2 cache. MCDRAM is integrated on-package while DRAM is off-package and connected by six DDR4 channels.

3) We show that applications with regular memory access patterns largely benefit the HBM. On the contrary, applications with random access patterns do not benefit HBM.
4) We identify the HBM higher latency as the main obstacle to achieve performance gain in applications with random memory access.
5) We show that the use of multiple hardware threads in the same core is critical to take advantage of HBM high bandwidth.
6) We provide a guideline for setting correct expectation for performance improvement on systems with 3D-stacked high-bandwidth memories.

The rest of this paper is organized as follows: Section II provides a brief description of the Intel KNL memory system. Section III describes our system set-up, software environment and selected benchmarks and mini-applications. We show our experimental results in Section IV and compare our approach and methodology to previous work in Section V. Finally, we discuss the results and conclude the paper in Section VI.

## II. Intel Knights Landing Processor

In this section we briefly summarize the main characteristics of the Intel KNL processor, with particular emphasis on the memory sub-system and its configuration options. A detailed description of the system can be found in the literature [9].

Fig. 1 presents a KNL processor and its connection to the main memory. Each computing core has a private 32KB L1 cache (not presented in the diagram) and supports four hardware threads. Computing cores are grouped into tiles connected through a *mesh* network on-chip. Each tile consists of the two computing cores that share a 1-MB L2 cache. The L2 cache of these tiles are maintained coherent by a distributed tag directory implemented with a MESIF protocol, which enables cache-to-cache line forwarding. The example in Fig. 1 shows that the 36 tiles provide a total of 36 MB L2 cache.

In this work, we are mainly interested in the KNL memory sub-system. Computing cores in a KNL system are connected to two different memory technologies: MCDRAM and DDR. The MCDRAM modules are located on package while DDR is off-package and connected by six DDR4 channels driven by two memory controllers. The 3D architecture and the shorter distance to the computing cores allow MCDRAM to deliver a peak bandwidth of ~400 GB/s while DDR can deliver ~90 GB/s [9]. This 4× difference in bandwidth can be crucial for many applications that are memory bound. However, MCDRAM is also more expensive and has much smaller capacity compared to DDR. On the current version of KNL, MCDRAM is extensible up to 16 GB while DDR can reach 384 GB in size. An optimal usage of KNL processor needs to consider the features of both memory technologies, aiming for a balance of the two.

MCDRAM can be configured in three modes: *cache, flat or hybrid* mode.

**Cache**: The default mode is cache mode, where MCDRAM is used as the last-level cache (i.e., L3 cache) for DDR transfers. In this mode, the MCDRAM is completely managed by the hardware so that it is transparent to users and OS. This mode requires no code modifications or manual allocation of data structures. Thus, it is the most convenient mode in terms of manual efforts and flexibility. The MCDRAM cache protocol uses a direct mapping scheme, which is an effective protocol but can result in higher conflict misses. In this mode, the memory hierarchy consists of L1 and L2 SRAM caches, an L3 cache backed by MCDRAM and a homogeneous main memory backed by DDR.

**Flat**: When MCDRAM is set in the flat mode, it works side-by-side with DDR. Together, they form a heterogeneous main memory. This mode requires either code modifications or using `numactl` to explicitly control the data placement. If an application can isolate data structures that benefit from the higher bandwidth of MCDRAM, it is feasible to have fine-grained data placement using heap memory management libraries, such as the memkind library [10]. On the other hand, if the entire application fits into MCDRAM, coarse-grained placement by `numactl` is more convenient. In flat mode, the memory hierarchy includes L1 and L2 caches and a heterogeneous main memory composed of small capacity MCDRAM and large capacity DDR.

**Hybrid**: The last memory mode is the hybrid mode. In this mode, MCDRAM can be partitioned into two parts, one in cache mode and the other in flat mode. This mode should be used when the data structures in an application are accessed in different patterns. Changing partition requires a system reboot and modification of the BIOS to indicate the ratio of MCDRAM in cache mode and the one in flat mode. Thus, it is arguably cumbersome to change the MCDRAM partitioning for each application.

Table I: List of Evaluated Applications

| Application | Type | Access Pattern | Max. Scale |
|---|---|---|---|
| DGEMM | Scientific | Sequential | 24 GB |
| MiniFE | Scientific | Sequential | 30 GB |
| GUPS | Data analytics | Random | 32 GB |
| Graph500 | Data analytics | Random | 35 GB |
| XSBench | Scientific | Random | 90 GB |

## III. Experimental Environment

This section describes the hardware and software setup used in this work and provides a brief description of the tested applications.

### A. Hardware Setup

We carry out experiments on a cluster of 12 KNL-based compute nodes on the Cray Archer supercomputer [11]. Each node is equipped with a 64-core KNL processor (model 7210) running at 1.3 GHz. Each core can support up to four hardware threads. The testbed is configured with 96-GB DDR and 16-GB MCDRAM. MCDRAM consists of eight 2-GB modules. The six DDR4 channels are running at 2.1 GHz. Two compute nodes are pre-configured with MCDRAM in flat mode, while the remaining ten nodes have MCDRAM configured in cache mode. All nodes use the quadrant cluster mode for the mesh of tiles. These compute nodes are connected via Cray's proprietary Aries interconnect.

### B. Benchmarks and Applications

We select benchmarks and applications whose performance is sensitive to the characteristics of the memory system. The list of applications is presented in Table I. The applications are summarized as follows:

- **DGEMM** [12] benchmark performs dense-matrix multiplication.The memory access pattern is sequential and optimization in data locality is crucial. For this evaluation, we compile the code linking to the Intel Math Kernel Library (MKL), which provides highly-optimized multithreaded kernels. We report Giga floating point operation per second (GFLOPS) for performance evaluation.
- **MiniFE** [13] is a DOE proxy application that is representative of implicit Finite Element applications. The most performance critical part of the application solves the linear-system using a Conjugate-Gradient algorithm similar to HPCG benchmark. We report the total Mflops in the CG part of the application output for performance evaluation.

- **GUPS** [14] is a synthetic benchmark that measures the Giga-updates-per-second (GUPS) by reading and updating uniformly distributed random addresses in a table. The memory access pattern is random with poor data locality. This synthetic problem is often used for profiling the memory architecture. We report GUPS measured by the benchmark for performance evaluation.
- **Graph500** [15] represents data-analytics workloads. The memory access pattern is data-driven with poor temporal and spatial locality. Thus this application is featured with random access pattern. We use the reference implementation version 2.1.4 from the official website( www.graph500.org/) for experiments. We use the OpenMP implementation using CSR compression format. We report the harmonic mean of traversed edge per second (TEPS) for performance evaluation.
- **XSBench** [16] mimics the most time consuming part of OpenMC, a Monte Carlo neutron transport code. XSBench isolates the macroscopic cross section lookup kernel. As the random input conditions for the macroscopic cross section lookups is retained in the application, the memory access is random and demands for large memory capacity. We use a large fixed problem-size and vary the number of grid points (option -g) to scale up the memory footprint of the test. We report the rate of performed lookups per second (lookups/s) for performance evaluation.

All applications are compiled using the Intel compiler version 17.0.0.098 (gcc version 4.8.0 compatibility) with OpenMP, `-xMIC-AVX512` flags and optimization option `-O3`. Applications are linked to the Cray MPICH library version 7.4.4.

### C. Execution Setup

For all applications, we scale up the problem size to the maximum containable size for a single node. We use OpenMP threads to take advantage of all the 64 computing cores. By default, 64 OpenMP threads are used for experiments unless otherwise specified.

We consider three memory configurations in the scope of this paper. First, if HBM is configured in flat mode and all memory allocation is bound to HBM, we refer this set-up as *HBM* in the results as it only uses the high-bandwidth memory. Second, if HBM is configured in flat mode but all memory allocation is bound to DRAM, we refer this setup as *DRAM* in the results. Third, if HBM is configured in cache mode, this configuration uses DRAM as main memory and HBM as the last-level cache. We refer this setup as *Cache Mode*.

We use `numactl` to control the data placement into the two memories. When HBM is in flat mode, there are two NUMA nodes available, corresponding to DRAM node and HBM node respectively. When HBM is in cache mode, there is only one NUMA node available. We report the NUMA domain distances obtained by the command `numactl --hardware` in these two setups in Table II.

Table II: The left panel presents the distance between DRAM node (0) and HBM node (1) when HBM is flat mode. The right panel shows that only one NUMA node is available when HBM is in cache mode.

| Distances: | 0 (96 GB) | 1 (16 GB) |
|---|---|---|
| 0 | 10 | 31 |
| 1 | 31 | 10 |

| Distances: | 0 (96 GB) |
|---|---|
| 0 | 10 |

The *DRAM* configuration uses HBM in flat mode and use `numactl --membind=0` to allocate all data in DRAM. The *HBM* configuration uses HBM in flat mode and use `numactl --membind=1` to allocate all data in HBM. The *Cache Mode* configuration uses HBM in cache mode and use `numactl --membind=0` for consistency even though there is only one NUMA domain available.

## IV. Evaluation

In this section, we evaluate the performance benefit from HBM in a hybrid memory system. HBM can be used as main memory or as the cache for DRAM transfers. Thus, for each application, we carried out two identical experiment sets, one in each configuration. To establish the baseline, we also repeat each set of experiments on DRAM. We first use two micro-benchmarks to understand the hardware characteristics of the system. We then evaluate the performance impact from the memory sub-system to five applications that are representative for traditional scientific applications and data-analytics workloads.

### A. Memory Bandwidth and Latency

We first use micro-benchmarks to experimentally measure the peak bandwidth and latency of the testbed. The peak bandwidth is measured using the OpenMP version of STREAM benchmark with varying data sizes [17]. We report the results from the *triad* computational kernel in Fig. 2.

In this experiments we run one thread per core. Fig. 2 shows that DRAM achieves a maximum of 77 GB/s bandwidth (red line) while HBM reaches a maximum of 330 GB/s (blue line). Thus HBM provides a $4\times$ higher bandwidth than DRAM. As we will show later, HBM can reach as high as 420 GB/s using more hardware threads per core. The bandwidth that is measured in the cache mode configuration (black line) largely depends on the data size. In this configuration, we measure a peak performance of 260 GB/s when the input set size that is approximately half of the HBM capacity. Beyond that size, the measured bandwidth rapidly drops to 125 GB/s at 11.4 GB. When the input set is 22.8 GB, the cache mode performance even becomes lower than DRAM. The results indicate that HBM can provide the highest bandwidth if the data size fits in HBM. Cache mode is arguably the easiest way to use HBM and, although the caching protocol might introduces some overhead compared to direct data placement in HBM. The cache mode provides the second highest bandwidth and it is comparable to HBM up to 10GB. When the data size reaches the range that is larger than HBM size but still comparable to HBM (see 16-24 GB stream size range in Fig. 2), the cache mode provides higher bandwidth than DRAM. When the data size is much larger than HBM size, using HBM as cache for DRAM might even bring lower bandwidth than DRAM. This is likely due to the direct mapping scheme of HBM cache protocol, which results in higher capacity conflicts when data size increases. On the other hand, STREAM memory access pattern is very regular, thus the prefetcher can effective bring in data to the cache in advance, hiding memory latencies and reducing the differences between HBM in cache mode and DRAM. The STREAM benchmark also shows that the overhead of cache could even result in lower bandwidth than DRAM when the data size is larger than 24 GB.

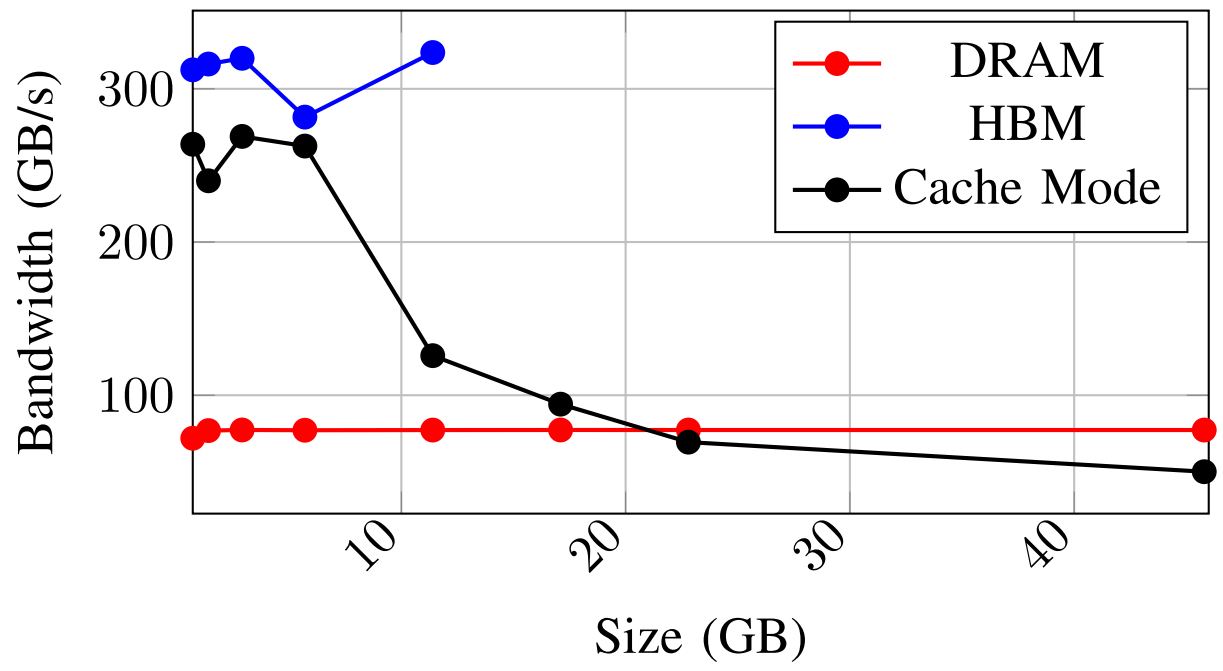

Figure 2: Peak bandwidth measured by the STREAM benchmark under the three memory configurations are reported. Measurements for HBM configuration stops when the data size does not fit into HBM.

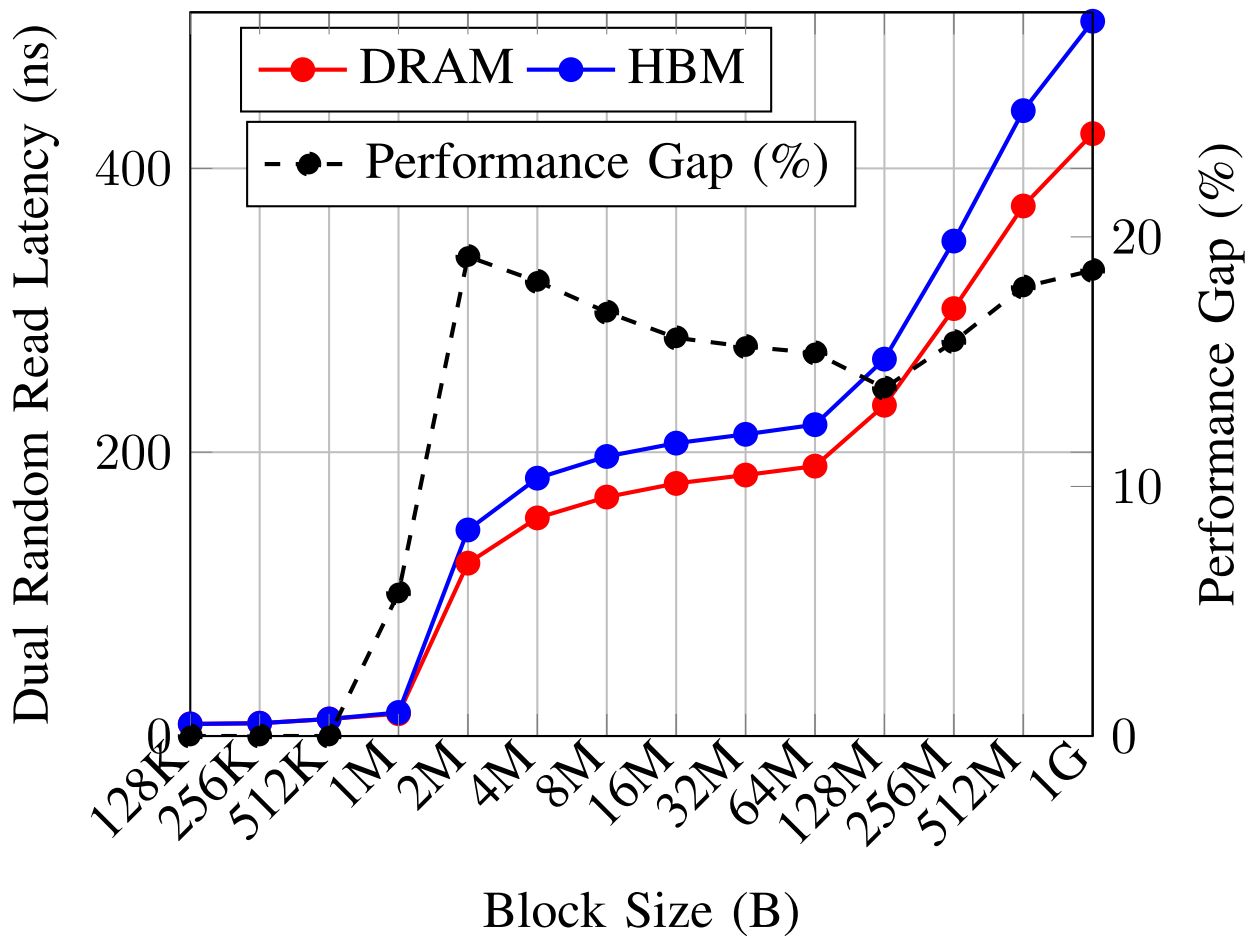

Figure 3: Dual random read latency to a buffer allocated in DRAM (blue) and HBM (red) respectively. As the buffer size increases, the latency includes effects from cache misses, TLB misses and page walk.

While HBM provides the highest bandwidth, it also presents a higher latency than DRAM Our results are consistent with previous work[18]: we report 154.0 ns latency for HBM and 130.4 ns for DRAM. This indicates that accessing to HBM could be $\sim 18\%$ slower than accessing DRAM. Besides the memory latency, we are also interested to quantify the correlation between the data size and time delay for accessing data at a random address. We use the open-source benchmark, TinyMemBench [19] to measure the read latency for random accesses to an allocated buffer with varying sizes. This benchmark reports the average measured latency with less than $0.1\%$ deviation. We report the measurement for performing two simultaneous random read (*dual read latency*) in Fig. 3. This measurement reflects the capability of handling concurrent requests by the underlying memory system. Thus it is more relevant for KNL out-of-order cores. We use the block size larger than L1 cache and also exclude L1 latency from the measurements.

Fig. 3 shows three tiers of the time delay in accessing random addresses. The first range is for block sizes smaller than 1 MB (local L2 cache). In this range, latencies are approximately 10 ns. The second range is block sizes that are smaller than 64 MB (two mesh L2 cache size). This range is characterized by a latency that is approximately 200 ns. In this region, the performance gap reaches the peak of 20% when the data size is just slightly larger than the tile local L2 cache. Beyond that, the gap starts decreasing to about 15% till the data size reaches four times of the mesh L2 cache, i.e., 128 MB. Starting from 128 MB, the latency values for DRAM and HBM increase with the increase of the block size. We observed that DRAM can be accessed 15-20% faster than HBM in the three block size ranges. The percentage difference between DRAM and HBM is represented with a black line in Fig. 3.

### *B. Impact of Access Pattern on Application Performance*

The top panels in Fig. 4 present the performance of applications with regular access patterns (DGEMM and MiniFE) with varying problem size. Binding memory allocation to HBM provides the best performance for both DGEMM and MiniFE. The two applications achieve $2\times$ and $3\times$ the performance improvement by allocating data structures in HBM compared DRAM, respectively. The bottom panels in Fig. 4 present the performance of applications with random memory access patterns (GUPS, Graph500 and XSBench). We note that HBM does not bring any performance improvement, neither in flat mode (red bars) nor in cache mode (grey bars). On the contrary, these applications achieve the highest performance by only allocating data to DRAM (blue bars).

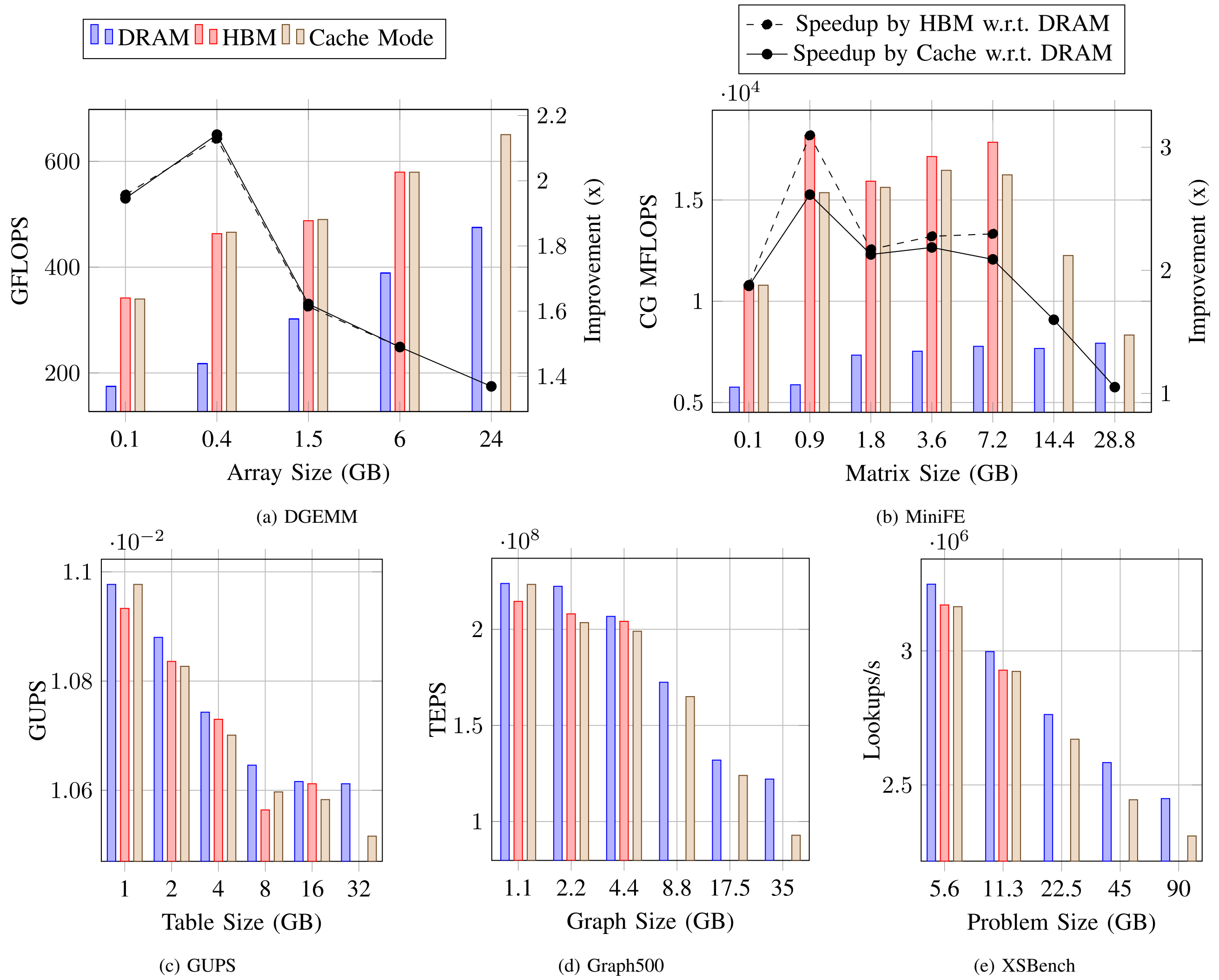

Figure 4: Application performance in different memory modes varying the problem size. Evaluation is based on single-node execution using 64 OpenMP threads. No measurements for HBM in flat mode (red bar) when the problem size exceeds its capacity.

By comparing the top and bottom panels of Fig. 4 it is evident that the application memory access pattern is an important factor to achieve high performance when using HBM. Regular applications are generally sensitive to memory bandwidth. These applications can directly benefit from HBM and show promising speedup. On the other hand, applications exhibiting random access pattern are more sensitive to memory latency, thus they are penalized by the higher latency of HBM. By Little's Law [20], the memory throughout equals to the ratio between the outstanding memory requests and the memory latency. If an application has regular access pattern, both prefetcher and the out-of-order core can perform well to increase the number of memory requests. In that case, the bottleneck becomes the maximum number of concurrent requests that can be supported by the hardware. Since HBM provides $4\times$ higher bandwidth than DRAM, applications with regular access benefit significantly from such high-bandwidth memories. In the case of random access pattern, the memory addresses cannot be predetermined and the number of requests remains small. Thus the throughput is bound by the latency. The benchmarking on the testbed shows that DRAM has about 20% lower latency than HBM (presented in Fig. 3 and also reported by [9]). As a result, the latency-bound applications do not benefit from HBM but performs better on DRAM.

In general, we observe the cache mode does not show clear advantage when compared to HBM or DRAM, but performance in this mode generally fall in between the highest and the lowest. For applications that benefit from HBM, the cache mode is an effective way to achieve high performance when the data size is larger than the HBM capacity.

### C. Impact of Problem Size on Application Performance

The problem size directly determines whether it is feasible or convenient to store the application data entirely in HBM. First, even if the application access pattern favors HBM in flat mode, the problem size might be too large to fit in HBM. Second, HBM can be used to augment the total memory capacity of the platform, which might be beneficial for applications with very large input sets, especially on system with a relatively small DRAM. On platforms with similar ratio between DRAM and HBM, the only way to run some large problems might be to use both HBM and DRAM side-by-side, e.g., setting HBM in flat mode and interleaving memory allocation between the two memories.

The problem size also impacts the application performance from hybrid memory system. When the problem size fits in HBM, the optimal configuration for applications with regular access is to use HBM in flat mode and allocate data only in HBM. This is clear from the top panel of Fig. 4, where the configuration with HBM (red bars) achieve the highest performance for both applications over all problem sizes that fit in HBM. When the problem size is larger than HBM but still comparable to its capacity, the cache mode can significantly improve the performance compared to the performance on DRAM. However, the performance improvement from the cache mode decreases as the problem size increases. In the case of the MiniFE, the improvement from the cache mode drops to $1.05\times$ when the problem size is nearly twice HBM capacity. This performance degradation in cache mode is consistent with the results of the STREAM benchmark. This result can be useful for determining multi-node configuration of a given problem. If the application has good parallel efficiency across multi-nodes, with enough compute nodes, the optimal setup is to decompose the problem so that each compute node is assigned with a sub-problem that has a size close to the HBM capacity.

The performance gap between DRAM and HBM increases as the problem size increases in latency-bound applications. The bottom panels of Fig. 4 show that when the problem is small, i.e., 1 GB table for GUPS, 1.1 GB graph for Graph500 and 5.6 GB grid points for XSBench, running on DRAM (blue bar) or HBM (red bar) or cache mode (grey bar) results in small performance difference. However, when the problem size is larger than the HBM capacity, the performance gap increases. For instance, running Graph500 on a large scale graph, allocating data only to DRAM leads to a $1.3\times$ performance increase compared with the performance from using HBM in cache mode. There are two main reasons for this result. First, as presented in Fig. 3, the difference between the random read latency to DRAM and to HBM increases when the data size increases. Second, these applications have poor data locality so that data brought in cache is not used. In this case, the cache brings limited benefit to hide memory latency but the overhead of cache is still paid. As HBM in cache mode uses direct mapping scheme, when the data size increases to be comparable to the HBM capacity, the number of capacity conflicts increases, further lowering the performance of cache mode.

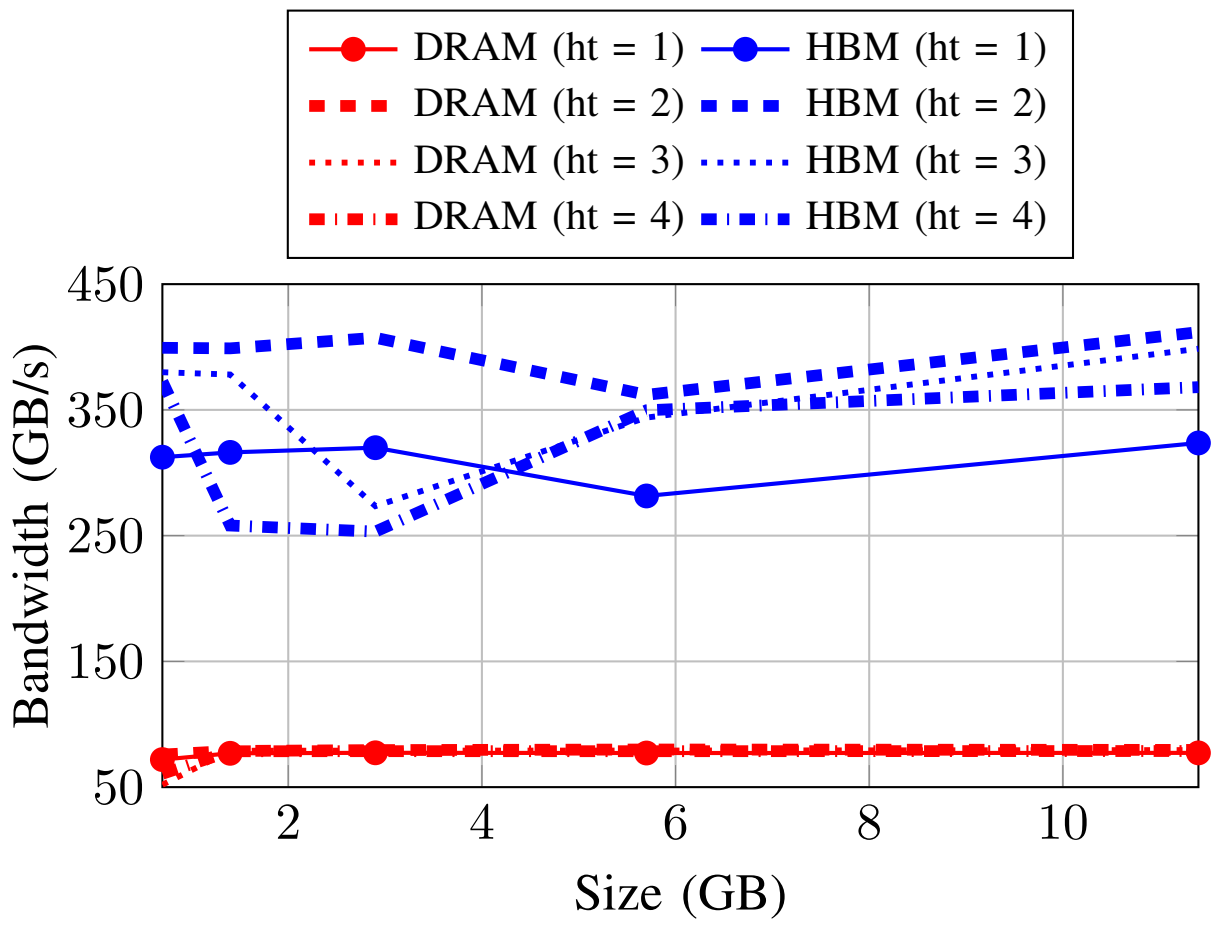

Figure 5: Impact on memory bandwidth by using hardware threads on DRAM and HBM measured by the STREAM benchmark.

### D. Impact of Hardware Threads on Application Performance

Each KNL core supports four hardware threads so that up to 256 hardware threads are available on a single node. Intel hyper-threading allows multiple hardware threads to issue instructions without performing a complete context switch. Hardware threads that are waiting for data or an internal resource are temporarily parked while the processor core fetches instructions from another hardware thread. Thus hardware multi-threading is generally considered as an effective way to hide latency, especially for latency-bound applications.

On Intel KNL, using additional hardware threads per core might help hide memory latency, which can results in an overall performance improvement especially when accessing HBM in flat mode. As the first step, we quantify the impact of using hardware threads on the memory bandwidth. We measure the aggregated bandwidth by using one to four hardware threads per core when running the STREAM benchmark presented in previous section. The measured bandwidth is reported in Fig. 5. For HBM, using two hardware threads per core (blue dashed lines) reaches $1.27\times$ the bandwidth measured with only one hardware thread per core at all data sizes. Other thread counts have varying performance on different sizes, but the graph shows that one thread per core cannot achieve peak bandwidth. Applications using DRAM (red lines) only slightly benefit from more hardware threads so that all four red lines overlap with each other in Fig. 5.

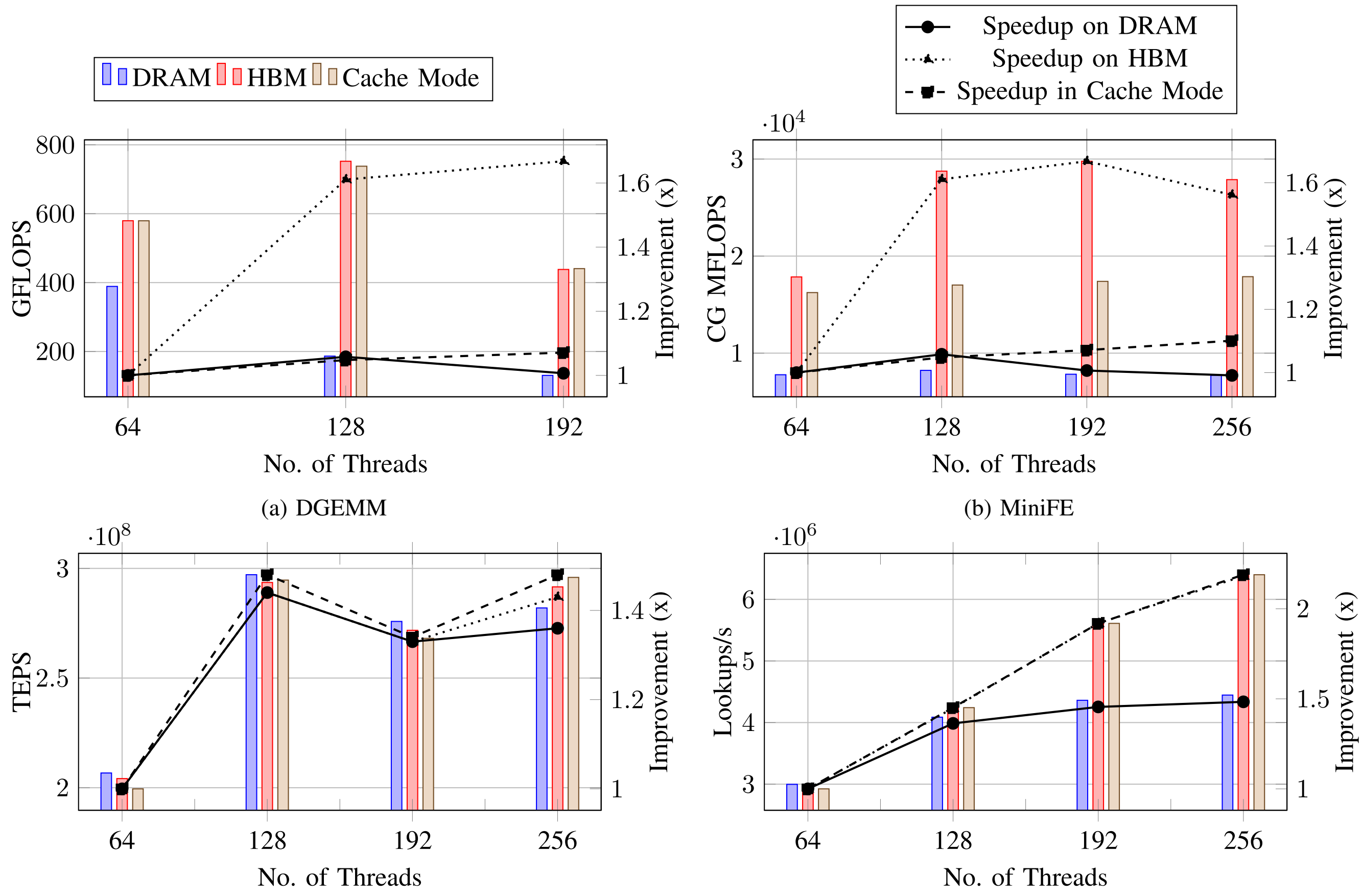

(a) DGEMM

(b) MiniFE

(c) Graph500

(d) XSBench

Figure 6: Impact on the performance of applications with sequential access (DGEMM and MiniFE in the top two panels) and on applications with random memory access (Graph500 and XSBench in the bottom two panels) using different number of hardware threads. The black lines indicate the speedup by using $X$ threads w.r.t 64 threads.

The top panel of Fig. 6 shows the performance of applications with regular memory access patterns, DGEMM and MiniFE, varying the number of hardware threads. As the plots show, these applications significantly benefit from HBM when using multiple hardware threads because they achieve higher aggregated bandwidth. We observe a $1.7\times$ performance improvement when moving from 64 (one hardware thread/core) to 192 (three hardware threads/core) hardware threads for both DGEMM and MiniFE.[1] As explained above, the additional hardware threads do not effectively hide DRAM latency, thus neither DGEMM nor MiniFE show particular performance benefit from using hyper-threading.

Hyper-threading can be used to hide memory latency in irregular applications as well. We quantify the impact of hyper-threading on Graph500 and XSBench performance. The results are reported in the bottom two panels of Fig. 6. Both applications benefit from additional hardware threads in all three memory configurations. XSBench reaches the highest performance ($2.5\times$) by using 256 threads in HBM and in cache mode, while the DRAM configuration achieves $1.5\times$ performance compared to the execution with one thread/core. For this application, multi-threading can effectively hide latency so that the best configuration with hyper-threading (HBM) is different from the best option with a single thread/core (DRAM). Graph500 shows similar improvements but DRAM still remains the best configuration, as it shows the highest performance when using 128 threads. All three memory configurations achieve $1.5\times$ performance when using multiple hardware threads. These results indicates memory-bound applications can significantly benefit from multi-threading, both in HBM and DRAM configurations.

Our results also show that the optimal number of hardware threads varies from application to application. XSBench shows a trend of higher performance from larger count of threads on all three memory configurations. In Graph500, all memory configurations achieves the best performance on 128 threads. Different memory configurations have different optimal counts of threads in MiniFE. An extensive study to find the optimal setup for each application is out of the scope of this work. However, it is important to note that HBM is more sensitive to multi-threading and it could become the optimal options even for latency-bound applications like XSBench.

[1]Note that results relative to DGEMM with 256 hardware threads are not available as the run can not complete successfully.

## V. Related Work

Next-generation supercomputers will likely feature a hybrid memory sub-systems consisting of conventional DRAM systems next to 3D-stacked memory and/or Non-Volatile RAM. 3D-stacked DRAM is a high density, high bandwidth memory obtained by stacking multiple DRAM semiconductor dies on top of each other and communicating with Through-Silicon-Vias (TSV) [21]. Currently, besides the Intel KNL analyzed in this work, other systems such as Altera Stratix 10 and Micron Hybrid Memory Cube are also equipped with HBM [22], [23], [24]. Chang et al. [25] show that the most significant performance benefit of 3D-Stack memory is the largely increased bandwidth with respect to conventional DRAM. However, the same study also shows that latency to HBM is not reduced as expected. One of the first examples of HBM is the MCDRAM that is a 3D-stacked DRAM that is used in the Intel Xeon Phi processor. Since the delivery of 3D-stack memories on the KNL, several performance evaluation with HPC applications have been conducted to understand the performance of the HBM. Rosales et al. [26] study the performance differences observed when using flat, cache and hybrid HBM configurations together with the effect of memory affinity and process pinning in Mantevo suite and NAS parallel benchmark. Barnes et al. [27] present the initial performance results of 20 NESAP scientific applications running on KNL nodes of the Cori system and comparing KNL hardware features with traditional Intel Xeon architectures. This study mainly targets at how to effectively run NESAP applications in the Cori system whereas we focus on giving general guidelines on what kind of applications characteristics benefit from running on a hybrid memory system. Moreover, both works analyze traditional scientific applications while we also evaluate the data-analytics applications. Other studies [5], [6] focus on the performance of a specific application on a KNL system and show how to optimize the application to increase the performance benefit. More specifically, Bálint et al. [6] shows of a performance analysis of linear solvers and kernels used in Lattice Quantumchromodynamics (QCD) on 16 KNL nodes. This work also emphasizes the importance of using hyper-threads to effectively exploit HBM bandwidth. DeTar et al. [5] optimize a staggered conjugate gradient solver in the MILC application for KNL and report that use of HBM brings a large improvement for the MILC application.

## VI. Discussion and Conclusions

The 3D-stacked high-bandwidth memories are emerging on mainstream processors and hardware accelerators. These new memory technologies potentially bring significant improvement to bandwidth-bound applications but they are currently more expensive and present an higher latency than traditional DRAM technologies. Thus their usage is limited to small capacities. To mitigate these limitations, current compute nodes often consist of a small capacity HBM (high bandwidth, high latency) side-by-side with large capacity DRAM (lower bandwidth, lower latency). The optimal usage of such hybrid memory systems is still largely unknown, as the first hardware prototypes have just arrived to the market.

In this paper, we analyzed one of such hybrid memory systems, the Intel KNL, which features a 16 GB HBM next to a 96 GB DRAM. We studied a variety of traditional HPC and emerging data-analytics applications and identified memory access pattern, problem size, and threading as the main applications characteristics to determine whether HBM is beneficial or not. Our conclusions are backed up by experiments on real state-of-the-art compute node equipped with Intel KNL processors.

Our results show that applications with sequential access pattern are usually bandwidth-bound and benefit the most from HBM. Our evaluation indicates that using HBM provides up to $3\times$ performance improvement compared to storing data structures in DRAM. Applications with random access pattern, such as XSBench and Graph500, are instead latency-bound. Thus they suffer from the higher latency on HBM and achieve better performance when storing data structures in DRAM. For those applications, using multiple hardware threads per core can effectively hide HBM higher latency. Our results show that when using multiple hardware threads per core, HBM becomes a more effective solution. However, other applications, such as Graph500, might not be able to completely hide the memory latency, hence DRAM still gives the best performance.

Although we studied a specific hybrid memory system, we believe that our conclusions can be generalized to other heterogeneous memory systems with similar characteristics. Our study provides guidelines for selecting suitable memory allocation based on application characteristic and problem to solve. It also helps set the correct expectation of performance improvement when porting and optimizing applications on such systems. However, we used a coarse-grained approach, in which all application data structures are either stored in HBM or DRAM. In the future, we plan to investigate a finer-grained approach in which we can apply our conclusions to individual data structures and eventually employ Intel KNL hybrid HBM mode whenever necessary.

## Acknowledgments

This work was funded by the European Commission through the SAGE project (grant agreement no. 671500, http://www.sagestorage.eu/). This work used the ARCHER UK National Supercomputing Service (http://www.archer.ac.uk. This work was supported by the DOE Office of Science, Advanced Scientific Computing Research, under the ARGO project (award number 66150) and the CENATE project (award number 64386).